\documentclass[journal]{IEEEtran}
\IEEEoverridecommandlockouts
\usepackage{cite}
\usepackage{amsmath,amssymb,amsfonts}
\usepackage{algorithmic}
\usepackage{graphicx}
\usepackage{textcomp}
\usepackage{xcolor}
\usepackage{siunitx}
\usepackage{enumitem}
\usepackage[ruled]{algorithm2e}
\usepackage{subcaption}

\def\BibTeX{{\rm B\kern-.05em{\sc i\kern-.025em b}\kern-.08em
		T\kern-.1667em\lower.7ex\hbox{E}\kern-.125emX}}

\begin{document}
	
\title{Automated feature extraction and selection for data-driven models of rapid battery capacity fade and end of life\\

\thanks{This research was funded by EPSRC and Siemens. The authors are also grateful to Antti Aitio for providing comments on the manuscript before submission.}
}

\author{\IEEEauthorblockN{Samuel Greenbank and}
	\and
	\IEEEauthorblockN{David Howey, \textit{Senior Member, IEEE}}
\thanks{Samuel Greenbank and David Howey are from the Battery Intelligence Lab, Department of Engineering Science, University of Oxford, United Kingdom. \textit{Corresponding author: david.howey@eng.ox.ac.uk}.}
}

\maketitle

\begin{abstract}
Lithium-ion cells may experience rapid degradation in later life, especially with more extreme usage protocols. The onset of rapid degradation is called the `knee point', and forecasting it is important for the safe and economically viable use for batteries. We propose a data-driven method that uses automated feature selection to produce inputs for a Gaussian process regression model that estimates changes in battery health, from which the entire capacity fade trajectory, knee point and end of life may be predicted. The feature selection procedure flexibly adapts to varying inputs and prioritises those that impact degradation. For the datasets considered, it was found that calendar time and time spent in specific voltage regions had a strong impact on degradation rate. The approach produced median root mean square errors on capacity estimates under 1\%, and also produced median knee point and end of life prediction errors of 2.6\% and 1.3\% respectively.
\end{abstract}


\section{Introduction}
Predicting lithium-ion battery degradation during design and operation is a significant challenge, and a large number of techniques for this have been proposed in literature \cite{birkl2017degradationdiagnostics,li2019lifetimereview}. Machine learning models have recently been applied for forecasting battery state of health, but they remain limited by lack of transparency, and require careful choice of inputs \cite{li2019lifetimereview}. Battery degradation is typically measured using capacity fade or resistance increase. Degradation is caused by many mechanisms \cite{reniers2019modelsreview,dong2020RUL_brownianmotion}, and these may interact in various ways. Degradation mechanisms are influenced by a wide variety of factors, such as calendar time, high power use, low temperature use, and combinations of these \cite{birkl2017degradationdiagnostics}. However, a battery end-user can only measure time, current, terminal voltage and cell surface temperature, at best, plus very occasional capacity or resistance through bespoke characterisation tests if they are possible. Consequently, prioritising the mechanisms that drive ageing for a given battery and use case is challenging  \cite{birkl2017degradationdiagnostics}. For data-driven approaches, this challenge demands careful and flexible feature extraction.

There have been attempts at battery health prognosis using neural networks \cite{goebel2010RNNlifetime,wu2016FFNNlifetime}, support vector machines \cite{fermin2020kneepoints,patil2015SVMRUL,zhao2015SVMRUL} and Gaussian process regression \cite{richardson2017GPRhealth,richardson2019GPRtransition,goebel2008GPRRUL,liu2020ARDGPRcalendaraging,li2019GPRRUL,tagade2020GPRdeep,hu2020GPRRUL,lucu2020GPRpartB,zheng2019GPRselection}. Some previous researchers have made use of an open source dataset from the NASA AMES research centre \cite{richardson2017GPRhealth,richardson2019GPRtransition,li2019GPRRUL,tagade2020GPRdeep}, and much of this data shows an approximately constant degradation rate. However, lithium-ion cells have sometimes been shown to suffer from the onset of more rapid capacity fade or resistance increase later in life \cite{severson2019data1,baumhoefer2014productionvariability,attia2020data2}. This sudden acceleration in ageing is often referred to as a `knee point' \cite{baumhoefer2014productionvariability}. The knee point indicates that significant deterioration has occurred, and manufacturers may wish to replace cells at this point. Beyond the knee point, degradation accelerates rapidly and a cell can be considered to have lost its value for a given application, and there may also be safety concerns \cite{fermin2020kneepoints}. Forecasting it is therefore crucial for understanding the lifetime value of lithium-ion batteries \cite{attia2020data2}. Previous attempts at data-driven health prediction have tried to estimate the timing of the knee point \cite{fermin2020kneepoints,diao2019kneepoint} or the cycle life to 80\% capacity \cite{severson2019data1} respectively. In both cases these were point estimates, rather than predictions based on the full trajectory of health estimates. 

Forecasting battery state of health using machine learning approaches usually requires an assumption about the way the batteries are charged and discharged. Often, usage is implicitly considered to be fixed over time, with identical train and test use cases \cite{richardson2017GPRhealth}. However, changes in usage can be accommodated, for example by dividing usage into fixed sections of time and summing the health impacts piecewise over the sections. This approach has been successfully applied to capacity forecasting with a Gaussian process regression model \cite{richardson2019GPRtransition}. In this case, for the dataset used, a simple manual feature selection exercise identified calendar time and charge throughput as dominant inputs affecting capacity fade, although other features such as time periods spent in extreme current and temperature ranges could also be relevant for other datasets where batteries are used more intensively \cite{richardson2019GPRtransition}. 

For any machine learning  model, the input features decide the predictive performance. Several recent battery health publications have used a manually selected small feature set \cite{hu2020GPRRUL,lucu2020GPRpartB,zheng2019GPRselection,hu2020GPRNNFeatSel}. There is scope to automate the feature extraction/selection process, and test the model predictions over more general use cases.

\begin{figure*}[t]
	\centering
	\includegraphics[width=\textwidth]{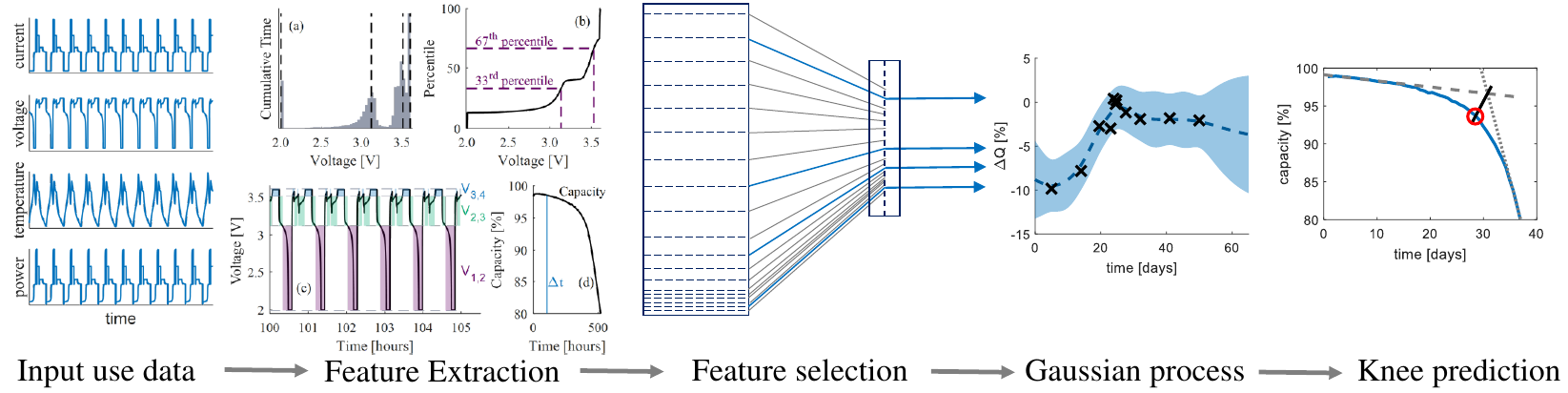}
	\caption{Workflow of the automated process proposed in this paper.}
	\label{fig:basic_premise}
\end{figure*}

We propose a novel automated feature selection approach, illustrated in Fig.\ \ref{fig:basic_premise}. The method extracts features from the data,  selects a relevant subset of key features, feeds their values into a Gaussian process regression model, and uses the output to produce an estimate of the capacity fade trajectory. The novel contributions of this work, are as follows: First, the feature extraction and selection process is automated and does not require users to choose beforehand what features they might expect to drive degradation. It is also transparent and easy to interpret. Second, the algorithm automatically chooses features that have reduced correlation with one another, avoiding unnecessary overlap. Third, it is fast and therefore able to deal with large datasets---it scales linearly with the number of input rows and quadratically with the number of features. Fourth, the approach can handle changing battery usage profiles - it does not assume cycling is always the same. Finally, we present (section \ref{sec:trialsetup}) comprehensive error metrics using a wide range of test conditions in order to demonstrate the robustness of the approach.

\section{Data sources}
Open source battery cycling (voltage, current, temperature) and capacity fade data were used for this work \cite{severson2019data1, attia2020data2}. The first dataset \cite{severson2019data1} consists of 135 lithium iron phosphate/graphite 18650 Li-ion cells (A123) that were cycled in a temperature chamber set at \SI{30}{\celsius}. All the cells underwent identical discharge cycles at 4C but had varied fast charging protocols from 3C to 8C \cite{severson2019data1}.
	
\begin{figure}[h!]
	\centering
	\includegraphics[width=.95\columnwidth]{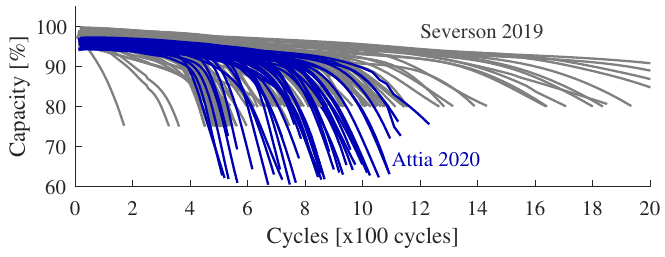}
	\caption{The data used in references \cite{severson2019data1} and \cite{attia2020data2}}
	\label{fig:stanford_capacities}
\end{figure}
	
The second dataset \cite{attia2020data2}, a follow up to the first, contains 45 cells which were cycled to failure, defined as 80\% of the nominal capacity of $1.1$ Ah. These cells were the same chemistry, size and manufacturer as those in the first dataset, and were tested at the same temperature setpoint and discharge rate as the previous test, but with a fixed 10-minute charging protocol \cite{attia2020data2}, using a wide range of C-rates from 3C to 8C. Both datasets were non-uniformly sampled with the lowest sampling rate being 0.2 Hz. In our work, the datasets were cleaned before use by removing cells with obvious experimental errors (as identified by the original authors of the dataset), and a shorter selection of the remaining data, with lifetimes between 15 and 40 days, was chosen. After this, 147 cells worth of data remained available.
	
\section{Methods}
Fig.\ \ref{fig:basic_premise} shows the pipeline from raw measured data through to feature extraction and selection, modelling, and finally capacity forecasting and knee point prediction. Feature extraction and selection are not dependent on battery chemistry, usage or history, therefore this approach should be able to handle a wide variety of different cells in the same manner. Since this is a supervised learning method, regular measurements of state of health (from check up tests) are required. The state of health metric used throughout this paper is the discharge capacity $Q$ \cite{severson2019data1, attia2020data2}, but other metrics (such as resistance) could instead be used if desired. Gaussian process regression was used to model the relationship between features and the \textit{change} in battery health $\Delta Q = Q(t+ \Delta t) - Q(t)$, where $Q(t+ \Delta t)$ is the measured capacity at time $t+\Delta t$ and $Q(t)$ is the measured capacity at time $t$. In other words, data was divided into discrete sections, each of time period $\Delta t$, each section having a single row of features and a single capacity change $\Delta Q$. The full capacity fade trajectory, $Q(t)$, for a given usage condition can then be constructed by summing all $\Delta Q$ over time, in a piecewise linear fashion.
	
\subsection{Feature Extraction}\label{sec:featprod}
The aim of the feature extraction and selection process is to reduce the dimensionality of the input data by producing metrics that represent the most important aspects affecting state of health \cite{deng2020SoH_Estimation_FeatSel_VoltageCurves}. For example if the raw input data has size $\mathbb{R}^{n\times m}$, where $n$ is the number of time points and $m$ the number of raw data streams, then we aim to have a feature set size $\mathbb{R}^{p\times q}$, where $p\ll n$. (In our case $q$ and $m$ are similar sizes, both relatively small.)

The frequency at which to extract features is a subjective choice, and involves a trade-off between computational complexity versus skipping useful information. If battery capacity is only measured occasionally (e.g.\ monthly), it would make sense to calculate features using data between capacity measurements, over some time period $\Delta t$. Here, the cells were cycled continuously at high rates, resulting in up to 28 cycles per day, and the cycling data was directly used for capacity measurements. We therefore could have chosen quite short intervals for $\Delta t$, we investigated the impact of different frequencies of feature extraction and found that using a very short interval ($\Delta t = 1$ hour) slowed down the computational speed significantly but did not change overall performance. Using a long interval ($\Delta t = 24$ hours) tended to miss important features such as rapid changes in ageing. Therefore, as a compromise, features were extracted for every $\Delta t = 12$ hours of data, equivalent to taking a health measurement every 9-19th cycle depending on charging protocol. Note, this is specific to the datasets used in this paper and may be different for other data.

Features can be extracted using any function of the raw data and there is a trade-off between complexity and simplicity \cite{murphy2012MLprobabilistic}. Here we created an automated process based on `time spent' in certain usage regions, based on the training data set. A pseudo-code representation of the feature extraction process is shown in Algorithm \ref{alg:feature_generation}. Time spent in different regions is a simple health indicator that can easily be understood and acted upon by a user. For the process to be automatic, the health indicators must be extracted based on data calculated from the training set. Hence feature extraction was split into two parts: 1) calculating the variable bounds, 2) extracting the feature data.

\begin{figure}
    \centering
    \includegraphics[width=.9\columnwidth]{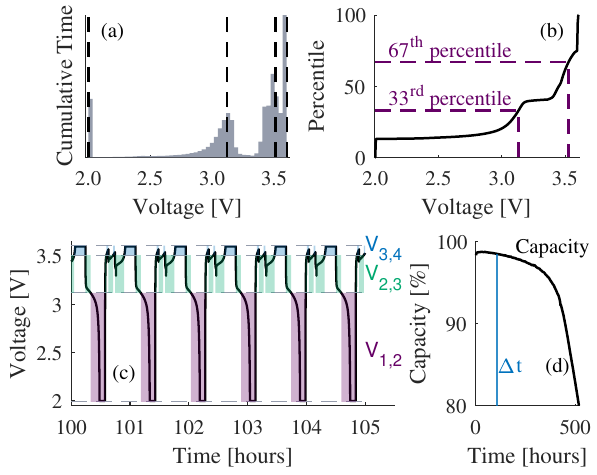}
    \caption{(a) Histogram and (b) Cumulative histogram, over all voltage values; (c) Example of three voltage features (coloured areas) overlaid on a subset of raw data; (d) Time interval of capacity data shown by (c)}
    \label{fig:example_feature_generation}
\end{figure}

Firstly, in order to create the variable bounds, time series of instantaneous absolute current $|I(t)|$, instantaneous power $P(t) = V(t)I(t)$ and instantaneous absolute power $|P(t)| = |V(t)I(t)|$ were calculated. Combined with the measured voltage, current and temperature time series, this gives $m=6$ basic input data streams. Then, for each of these input data streams, across the entire population of data available, a histogram was generated. Fig.\ \ref{fig:example_feature_generation}(a) shows an example of the histogram and \ref{fig:example_feature_generation}(b) the cumulative histogram for the voltage data.

\begin{table}
\begin{center}
\begin{tabular}{|c|c|c|c|c|}
	    \hline & Current & Voltage & Temperature & Power \\ 
		Percentile & [A] & [V] & [\SI{}{\celsius}] & [W] \\ \hline
		1\textsuperscript{st} & -4.00 & 2.00 & 30.0 & -12.84 \\
		33\textsuperscript{rd} & -0.53 & 3.12 & 32.8 & -1.08 \\
		67\textsuperscript{th} & 1.00 & 3.51 & 35.3 & 3.43 \\
		99\textsuperscript{th} & 6.00 & 3.60 & 40.3 & 21.34 \\ \hline
\end{tabular}

\caption{Example feature extraction variable bounds for a training set of 50 cells.}\label{tab:example_variable_bounds}
\end{center}
\end{table}

\begin{table}
\begin{center}
\begin{tabular}{|c|c|c|c|}
		\hline Number & Percentile limits & Label & Voltage Range \\ \hline
		1 & 1\textsuperscript{st} to 33\textsuperscript{rd} & V\textsubscript{1,2} & 2.00 V $< V <$ 3.12 V \\
		2 & 1\textsuperscript{st} to 67\textsuperscript{th} & V\textsubscript{1,3} & 2.00 V $< V <$ 3.51 V \\
		3 & 1\textsuperscript{st} to 99\textsuperscript{th} & V\textsubscript{1,4} & 2.00 V $< V <$ 3.60 V \\
		4 & 33\textsuperscript{rd} to 67\textsuperscript{th} & V\textsubscript{2,3} & 3.12 V $< V <$ 3.51 V \\
		5 & 33\textsuperscript{rd} to 99\textsuperscript{th} & V\textsubscript{2,4} & 3.12 V $< V <$ 3.60 V \\
		6 & 67\textsuperscript{th} to 99\textsuperscript{th} & V\textsubscript{3,4} & 3.51 V $< V <$ 3.60 V \\ \hline
\end{tabular}
\vspace{2pt}
\caption{Features extracted from voltage, $V$, profiles are proportions of time spent in specific ranges defined by the four chosen percentiles.}
\label{tab:example_features}
\end{center}
\end{table}

From the cumulative histograms, the values of each input corresponding to the 1\textsuperscript{st}, 33\textsuperscript{rd}, 67\textsuperscript{th} and 99\textsuperscript{th} percentiles were calculated, and each data series was divided into regions, as shown (again using voltage as an example) in Fig.\ \ref{fig:example_feature_generation}(b). Examples of these thresholds are given in Table \ref{tab:example_variable_bounds} and examples of feature types extracted using these thresholds are given in Table \ref{tab:example_features}. In addition to features extracted in this way, we also included time and the square root of time (both measured at the point capacity was measured) as features. 

The feature data was then calculated for each feature type and for every chunk of data of length $\Delta t$, corresponding to the time spent in each different region. This process resulted in the extraction of $q=74$ different types of features, and reduced $n=1.08\times10^8$ time points in the raw data across $m=6$ data streams, down to $p=7386$ rows of features. In section \ref{subsection:selection} we discuss how $q$ can be reduced substantially further to select only the most relevant $\approx5$ feature types. Typically, the most commonly selected feature was V\textsubscript{2,3}, the proportion of time spent between 3.12 V and 3.51 V, described in equation~\ref{eqn:gen_v23} and shown in grey in Fig.\ \ref{fig:example_feature_generation}(c).
\begin{align}
    \text{V\textsubscript{2,3}}(t_i) = \frac{\text{time between 3.12 V and 3.51 V}}{\Delta t = \text{time between $t_i$ and $t_{i-1}$}} \label{eqn:gen_v23}
\end{align}

The feature rows associated with the test sets were calculated in the same way, assuming perfect knowledge of the future cell use (but no knowledge of health, which is what the model predicts). Note that the cell usage in testing does not have to be and was not the same as the cell usage during training. 

A key advantage of our feature extraction approach over existing methods is that it models battery ageing flexibly as a function of different types of usage (for example, it will naturally account for users who leave the battery at high SOC for long periods of time, or do intensive cycling). Many existing approaches for health modelling are not able to do this because they use only features that we consider to be outputs rather than inputs to a model, in other words they capture instantaneous battery health metrics, but they do not necessarily capture the features that drive ageing. Examples of metrics used in this way include charge or discharge time between certain voltage limits \cite{zhao2015SVMRUL, hu2020GPRRUL, hu2020GPRNNFeatSel}, capacity \cite{hu2020GPRRUL, zheng2019GPRselection}, entropy \cite{hu2020GPRNNFeatSel}, resistance \cite{hu2020GPRRUL, goebel2008GPRRUL}, and incremental capacity analysis and differential voltage analysis based features \cite{fermin2020kneepoints, liu2020ARDGPRcalendaraging, severson2019data1, deng2020SoH_Estimation_FeatSel_VoltageCurves, hu2020GPRNNFeatSel}. Some papers have used charge throughput as an input feature \cite{richardson2019GPRtransition, lucu2020GPRpartB}, but we found in general this correlates well with time if the usage is consistent throughout life. Temperature-based features may also be of interest \cite{fermin2020kneepoints, lucu2020GPRpartB, hu2020GPRNNFeatSel}, although we have generally found voltage-based features to be more important in the datasets considered here.
	
\begin{algorithm}
\SetAlgoLined
\KwIn{1.\ current, voltage, temperature time series}
\KwIn{2.\ health metric, measured every $N$ cycles}

 1.\ calculate absolute current, power, absolute power\;
 2.\ assemble data matrix $\left[I \; V\;  T\;  |I|\;  P \; |P|\right]$\;
 3.\ \ForEach{column}{
 calculate histogram and cumulative histogram;\\
 calculate  1\textsuperscript{st}, 33\textsuperscript{rd}, 67\textsuperscript{th}, 99\textsuperscript{th} percentiles\;
 }
 3.\ \ForEach{variable}{
 \ForEach{percentile}{
 \textit{feature\_name} = \textit{variable}\textsubscript{start,end}\;
 }
 }
 4.\ \ForEach{battery time series}{
  i.\ record capacity measurement every 12 hours\;
  ii.\ record time and $\sqrt{\text{time}}$ at these points\;
  iii.\ \ForEach{feature\_name}{
  \ForEach{time interval $\Delta t$}{
  a.\ calculate time spent in range \textit{variable}\textsubscript{start,end}\;
  b.\ divide this by duration $\Delta t$\;
  }
  }
  
 }
 \KwOut{calculated feature data for all cells}
 \caption{Feature extraction and calculation} \label{alg:feature_generation}
\end{algorithm}

\subsection{Feature Selection}
\label{subsection:selection}
The aim of feature selection is to prioritise the types of features that are most important in affecting the battery health, and ignore the less important features. It is not expected that all feature types will affect the battery health equally, and redundant data or overfitting should be avoided. For example, previous work \cite{richardson2019GPRtransition} found that calendar time and charge throughput could be particularly significant for battery health prediction. 

Battery capacity fade trajectories are often quite smooth, which is unsurprising since loss of battery health is a cumulative process \cite{baumhoefer2014productionvariability,richardson2019GPRtransition}. For the cells used here the profiles can be split into just two or three distinct phases. It is therefore expected that only a small number of features might be required for modelling the state of health. 

Principle component analysis (PCA) is a common tool for dimensionality reduction \cite{murphy2012MLprobabilistic}. It was not used here because it would produce features which are linear combinations of inputs, whereas for simplicity, we wished to directly rank individual features, rather than combinations of features. 

\begin{figure}[h!]
	\centering
	\includegraphics[width=.65\columnwidth]{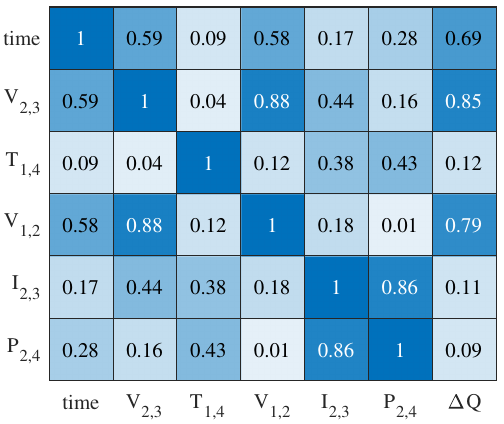}
	\caption{Example similarity matrix with 6 features and $\Delta Q$. Data was taken from a random subset of cells and feature labels use the scheme of Table \ref{tab:example_features}, with T, I and P as temperature, current and power respectively. This does not represent the full array of possible inputs used here.}
	\label{fig:simmateg}
\end{figure}

The feature selection method proposed here examines the similarity between features and changes in health. There are many methods available to measure similarity, such as correlation methods \cite{fermin2020kneepoints,zheng2019GPRselection,hu2020GPRNNFeatSel, deng2020SoH_Estimation_FeatSel_VoltageCurves, deng2020SoC_Estimation_FeatSel} and covariance functions \cite{murphy2012MLprobabilistic,rasmussen2006GPR}. We found most techniques had comparable performance, hence we chose to use the absolute value of Pearson's correlation coefficient, shown in equation \ref{eqn:pearsonsrank} for features $f_i$ and $f_k$. Here, \textit{cov} is covariance, $\sigma$ is standard deviation, and results are stored in similarity matrix $S$.
\begin{align}
	S_{i,j}(f_i,f_j) =& \left\|\frac{\text{\textit{cov}}(f_i,f_j)}{\sigma(f_i)\sigma(f_j)}\right\| \label{eqn:pearsonsrank} 
\end{align}
Elements of $S$ are bounded between 0 and 1, with higher values indicating stronger similarity. Figure \ref{fig:simmateg} gives an example of what a colourised version of $S$ might look like. 

Feature selection can now be performed; the simplest approach would be to select the features that correlate best with $\Delta Q$. However, that will likely lead to many redundant features being included. The health prediction model would likely perform equally with one of these features as with 10 of them, but with unnecessary increased computational complexity \cite{murphy2012MLprobabilistic,rasmussen2006GPR}. Therefore a step is included that removes redundant features, applying an upper limit on similarity across features. From the example of Fig.\ \ref{fig:simmateg}, one expects to select V\textsubscript{2,3}, then reject V\textsubscript{1,2} for being too similar, then select time as the second input feature. The impact of varying the maximum allowed shared correlation was investigated to verify its impact.
	
The selection process, which shares similarities with explicit orthogonalization \cite{lund2008orthogonalization} and simplifies wrapper-based methods for SoH estimation \cite{hu2020GPRNNFeatSel}, can be summarised as follows: (1) Find the feature correlating best with the change in capacity in the training set. (2) Remove all features which share a correlation coefficient greater than 0.85. (3) Repeat the previous steps until the required number of features is obtained. (The number of features required is chosen by the user, and is a trade-off between accu
racy and complexity.) In the example of Fig.\ \ref{fig:simmateg}, an appropriate selection might be features V\textsubscript{2,3}, V\textsubscript{1,2} and time, since these are similar to the required output $\Delta Q$. Features T\textsubscript{1,4}, I\textsubscript{2,3} and P\textsubscript{2,4} would be poor selections. 

\begin{figure}[h!]
	\centering
	\includegraphics[width=\columnwidth]{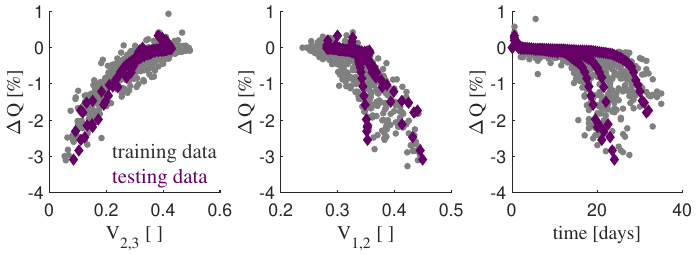}
	\caption{Example showing first three features selected for one training set plotted against change in capacity, $\Delta Q$. Purple test data overlaps grey training data which leads to successful forecasting.}
	\label{fig:featseleg}
\end{figure}

The effectiveness of the feature extraction and selection process may be demonstrated by considering how the features span the training and test datasets. If the process has been successful, there should be a strong overlap between train and test data as a function of battery health changes and the features being considered. An example of a successful outcome may be seen in Fig.\ \ref{fig:featseleg}. 

\subsection{Gaussian Process Regression}\label{sec:ml}

We selected Gaussian process (GP) regression to map from the selected input features to capacity transitions $\Delta Q$ since it is a flexible approach that makes very few assumptions on the function to be fitted \cite{richardson2017GPRhealth, richardson2019GPRtransition, liu2020ARDGPRcalendaraging, hu2020GPRNNFeatSel}. The dataset was split into training and test sections. The GP hyperparameters were fitted to the training data in the standard way using maximum marginal likelihood estimation, with Matlab's \texttt{fitrgp} function. The test datasets were used to quantify the model performance. A Mat\'ern 5/2 covariance function, shown previously to work well for this application \cite{rasmussen2006GPR,richardson2019GPRtransition},  was used. For two input data points, $x_i$ and $x_j$ the Mat\'ern 5/2 covariance is given by:
\begin{align}
r(x_i,x_j) =& \sqrt{\sum_k \frac{\left(x_{i,k} - x_{j,k}\right)}{\sigma_{l,k}^2}} \label{eqn:r_ard} \\
\kappa_{m52}(r) =& \sigma_f^2 \left(1 + \sqrt{5}r + \frac{5}{3}r^2\right)\exp\left(-\sqrt{5}r\right) \label{eqn:matern52}
\end{align}
The two hyperparameters $\sigma_f$ and $\sigma_l$ represent the magnitude and lengthscale of the covariance. The covariance function used automatic relevance determination, which allows the length-scale hyperparameter, $\sigma_l$, to have a different value for each input feature, $k$ \cite{liu2020ARDGPRcalendaraging}.

The test set inputs, unseen by the algorithm in terms of fitting, was used on the trained GP model to produce capacity transition test predictions and their associated $\pm 2\sigma$ credible intervals.

In summary, the feature selection process produced a set of inputs which correlated with the output $\Delta Q$. Together, the features formed an approximate linear model, which the GP then improved on to produce an accurate non-linear model of capacity fade.

\subsection{Knee Point Identification}

\begin{figure}[h!]
	\centering
	\includegraphics[width=.8\columnwidth]{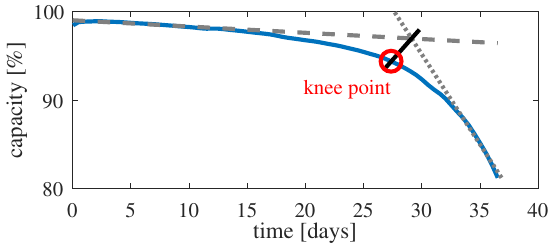}
	\caption{Knee point calculation.}
	\label{fig:knee_point_calc}
\end{figure}

The capacity transition forecasts were summed over time to form predicted capacity fade profiles, starting with the initial value of capacity which was assumed to be known. From this, various techniques can be used to locate the knee point position \cite{fermin2020kneepoints,diao2019kneepoint}. Here we chose to fit the early and late life capacity fade gradients using linear regression, then calculated their angle bisector and found the intersection of this with the capacity fade curve, see Fig.\ \ref{fig:knee_point_calc}.

\subsection{Evaluation Metrics}

The performance of the approach was evaluated by assessing the accuracy of the predicted capacity profiles of the test set using various metrics. The first is the root mean square error, RMSE\textsubscript{$Q$}, of the predictions. This is an effective metric for predictive performance, but it could be significantly influenced by a single, poor $\Delta Q$ prediction. Therefore a second metric to consider is the root mean square error of the transitions, RMSE\textsubscript{$\Delta Q$}, calculated the same way, but with only the capacity transition data.

The third capacity-based performance metric was the end of life percentage error, PE($t$\textsubscript{EoL})$=100( \hat{t}_\text{EoL}-t_\text{EoL})/t_\text{EoL}$, which measures the percentage difference in time between the predicted $\hat{t}_\text{EoL}$ and observed end of life, $t_\text{EoL}$.

The fourth metric is the knee point prediction accuracy, which may be evaluated in the same way as the end of life error but using the knee point position in time, PE($t$\textsubscript{EoL}).
 
For all evaluation metrics, the median value and 95\textsuperscript{th} percentile are reported. Mean errors are also quoted, for comparative purposes.

\subsection{Trial Setup}\label{sec:trialsetup}

The feature selection technique was tested in two ways: a large-scale test, and a limit testing experiment. The large-scale test quantified the general performance of the approach using the standard k-fold cross-validation technique \cite{murphy2012MLprobabilistic}, using 20 randomly sampled datasets, each with 100 training cells and 47 test cells. In the limited data test, a subset of the trial data was selected to investigate the impacts of missing data on performance, particularly late life data.

For the large-scale test, temperature, voltage and current data from 147 cells was used, with lifetimes ranging from 15 to 40 days, taken from 4 different batches \cite{severson2019data1, attia2020data2}. Health measurements were calculated for each cell every 12 hours from discharge capacity measurements, and smoothed using a moving average. In each trial, 30 test cells and 100 training cells were randomly selected. This whole process was repeated 20 times, giving 600 estimates of the knee point and end of life. Training and test datasets were completely separate. For the purpose of testing, the entire capacity trajectory was forecasted, from day 1, and the knee point and end of life was calculated from that.

\begin{figure}
		\centering
		\begin{subfigure}[b]{0.5\columnwidth}
			\includegraphics[width=\columnwidth]{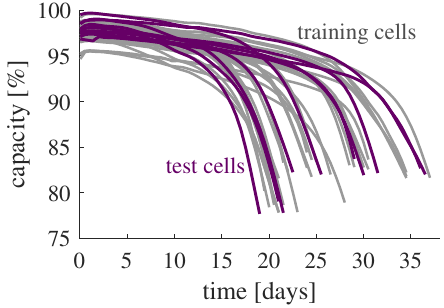}
			\caption{Capacity data used.}
			\label{fig:limit_test_cap_profiles}   
		\end{subfigure}             
		\begin{subfigure}[b]{0.36\columnwidth}
			\includegraphics[width=\columnwidth]{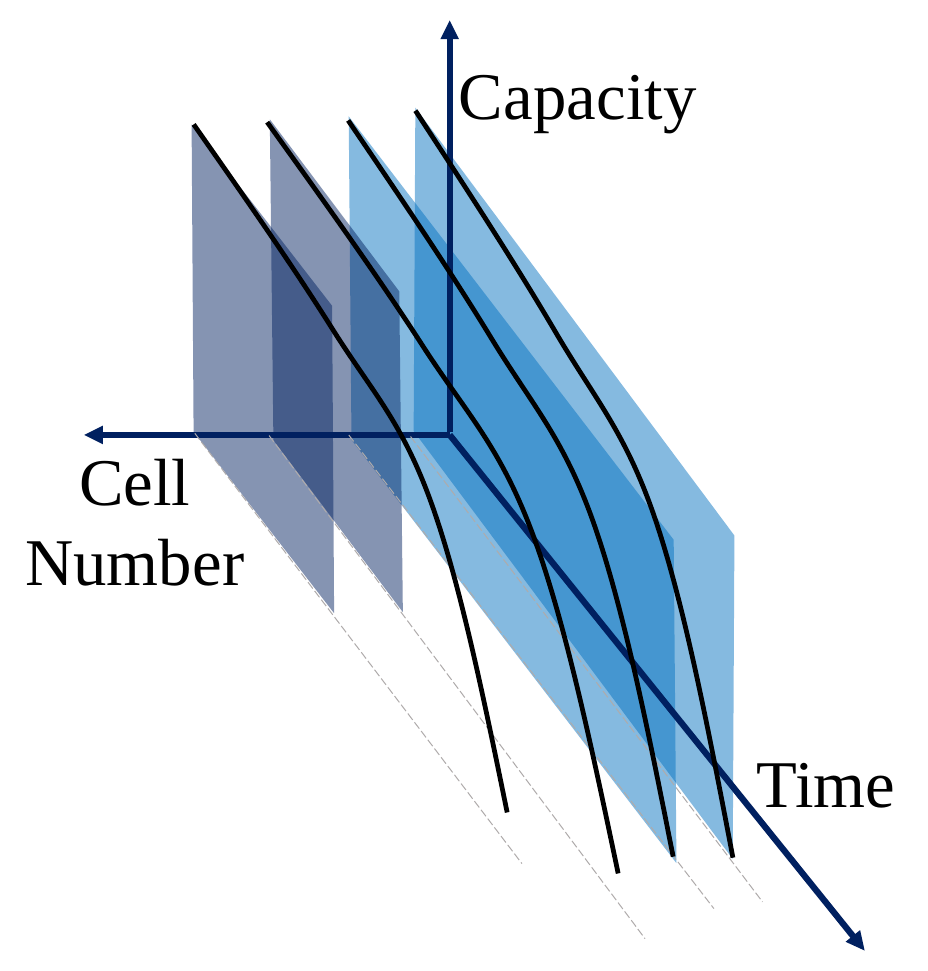}
			\caption{Example training set.}
			\label{fig:limited_data_diagram}
		\end{subfigure}             
		\caption{Trial set up for the limited late life data test. Capacity profiles shown with an example train/test split. The shaded regions in the training data show data is either taken from only early life (dark blue) or the full life (light blue).}
		\label{fig:limit_test}
\end{figure}

The limited late-life data test used a 40 cell subset of the data in reference \cite{severson2019data1}. In all instances, there were 10 testing cells and 30 training cells, with all cells and an example train/test splitting shown in Fig.\ \ref{fig:limit_test_cap_profiles}. The training sets were used to extract and select 5 features. However data was removed from a random set of cells, leaving between 3 and 30 cells with full life data, schematically shown in Fig.\ \ref{fig:limited_data_diagram}. This replicates a real-life scenario where limited full life data is available.

\begin{figure*}
	\centering
	\includegraphics[width=\textwidth]{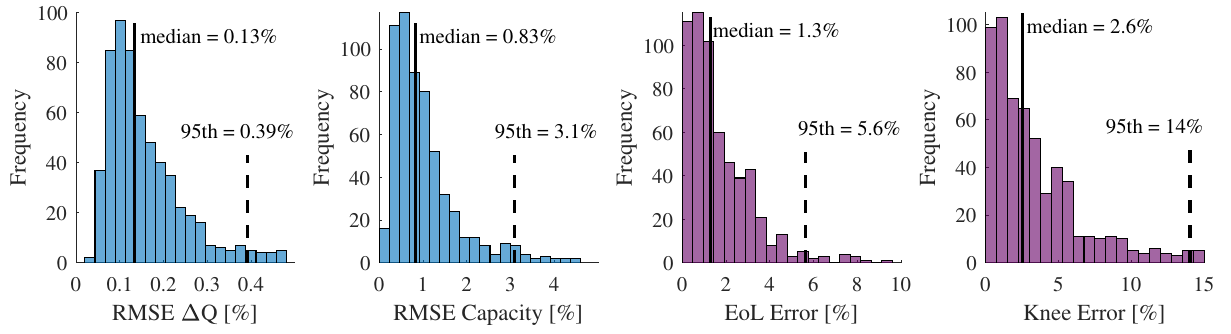}
	\caption{Histograms of the results for RMSE\textsubscript{$\Delta Q$}, RMSE\textsubscript{$Q$}, EoL percentage error and knee point percentage error, with medians and 95\textsuperscript{th} percentiles shown.}
	\label{fig:result_histogram_plots}
\end{figure*}

A short investigation into the impact of varying the maximum shared correlation among the input features was performed. For speed, it used a fixed set of features, but still randomly selected cells for the training and test sets. There were 20 repeats at each of 16 test points, ranging from 1 to 0.7, with 50 cells used for each of training and testing.

Finally, a further test was conducted looking at the relationship between the number of features and the predictive performance. Tests using 1, 2, 3, 4, 5 and 10 features were performed using the large-scale test process detailed above. Further to this, a test with only time is used as the input, and no other features, was used to create a baseline performance for comparison.

\section{Results}
\subsection{Feature prioritisation}

The most commonly selected features using the process and datasets described above were found to be V\textsubscript{2,3} and V\textsubscript{1,2}. V\textsubscript{2,3} was selected every time while V\textsubscript{1,2} was selected second in 19 trials and removed in the other. Next most commonly selected was calendar time which was selected 18 times.

The length-scale hyperparameters of the GP provide an estimate of how relevant the down-selected features are for predicting capacity transitions. For example, the calendar time feature typically returned a length-scale of around 7 days. Since the data covers a range of $0$ to $40$ days, this suggests that this input is relevant to capacity fade.

\subsection{Kneepoint and end of life predictions}

Fig.\ \ref{fig:result_histogram_plots} shows histograms of prediction accuracy against test set, using the metrics previously introduced. The majority of the root mean square errors on capacity were very small. Table \ref{tab:large_trial_results_percents} summarises these results numerically. The median value was 0.83\%, while 95\% of profiles returned RMSE\textsubscript{$Q$} $<$ 3.1\% capacity (continuous black lines, Fig.~\ref{fig:result_histogram_plots}). The reduction in RMSE frequency at very small values of RMSE\textsubscript{$Q$} (below 0.3\%) suggests that the approach avoids overfitting. The capacity transition forecasts exhibited a median RMSE of 0.13\% capacity. This, combined with the small RMSE\textsubscript{$Q$} results, leads to good performance for the knee point and end of life forecasting.

\begin{figure}[h!]
	\centering
	\includegraphics[width=.8\columnwidth]{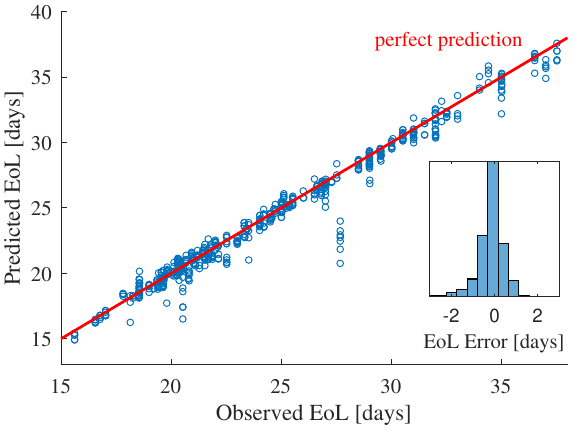}
	\caption{Scatter plot of observed end of life versus the prediction.}
	\label{fig:eol_results}
\end{figure}

Fig.\ \ref{fig:eol_results} shows a scatter plot of predicted verses observed end of life forecasts. The median absolute end of life forecast percentage error was 1.3\%, with a 95\textsuperscript{th} percentile at 5.6\% and a mean value of 2.0\%. 

Knee points were estimated from the capacity fade predictions and compared to the observations. The median absolute value of the knee point position error in time was 2.6\%, extending to 14\% once 95\% of the results are accounted for. 

\begin{table}
\begin{center}
	\begin{tabular}{|c|c|c|c|}
		\hline & Mean & Median & $95\%$ \\ \hline
		RMSE\textsubscript{$\Delta Q$} [\%] & 0.17 & 0.13 & 0.39 \\
		RMSE\textsubscript{$Q$} [\%] & 1.1 & 0.83 & 3.1 \\
		PE(\textit{t}\textsubscript{EoL}) [\%] & 2.0 & 1.3 & 5.6 \\
		PE(\textit{t}\textsubscript{knee}) [\%] & 4.2 & 2.6 & 14 \\ \hline
	\end{tabular}  
	\vspace{2pt}
	\caption{Summary results from the large scale trial}\label{tab:large_trial_results_percents}
\end{center}
\end{table}

Fig.\ \ref{fig:eol_results} suggests that there is a reasonably consistent performance across all lifetimes. Table \ref{tab:large_trial_results_days} presents the same results, but with respect to time. The knee point forecasts were less accurate than the end of life forecasts. This was caused by a slight over-prediction of late life gradient, impacting the knee point calculation but not significantly altering the end of life.

\begin{table}
\begin{center}
	\begin{tabular}{|c|c|c|c|}
		\hline Error & Mean & Median & 95\% \\ \hline
		EoL [days] & 0.49 & 0.29 & 1.5 \\
		Knee [days] & 0.74 & 0.45 & 2.48 \\ \hline
	\end{tabular}

	\caption{Summary results from the large scale test, in units of time (days)}\label{tab:large_trial_results_days}
\end{center}
\end{table}

\begin{table}
\begin{center}
\begin{tabular}{|c|c c|c c|c c|} 
    \hline
	& \multicolumn{2}{c|}{RMSE\textsubscript{$Q$} [\%]} & \multicolumn{2}{c|}{PE(\textit{t}\textsubscript{EoL}) [\%]} &
	\multicolumn{2}{c|}{PE(\textit{t}\textsubscript{knee}) [\%]} \\ 
	features & median & 95\textsuperscript{th} & median & 95\textsuperscript{th} & median & 95\textsuperscript{th} \\ \hline 
	10 & 0.54 & 2.6 & 0.92 & 4.5 & 1.7 & 13 \\ 
	5 & 0.83 & 3.1 & 1.3 & 5.6 & 2.6 & 14 \\
	4 & 0.95 & 3.2 & 1.3 & 6.2 & 3.2 & 13 \\
	3 & 0.96 & 3.1 & 1.5 & 6.2 & 3.6 & 13 \\
	2 & 1.3 & 5.1 & 1.6 & 9.0 & 5.1 & 16 \\
	1 & 1.8 & 5.0 & 2.1 & 11 & 7.2 & 17 \\ 
	time & 3.5 & 12 & 9.9 & 32 & 22 & 39 \\ \hline
\end{tabular}

\caption{Results varying the number of features (plus time). The bottom row is a trial using only time as an input.}\label{tab:results_vs_nfeats}
\end{center}
\end{table}

Adding more features as inputs tended to improve the predictive performance of the model in terms of the median error, Table \ref{tab:results_vs_nfeats}. However the 95\textsuperscript{th} percentile error did not significantly improve with more than 3 features plus time. Poor results were evident when only using time as the input; other features are also needed for accurate health prognosis.

To test the impact of lower sampling rate on the algorithm, we reduced the raw data to one data point every 100 seconds and found that the algorithm performance was unchanged.
	
\begin{figure}
    \centering
    \includegraphics[width=.9\columnwidth]{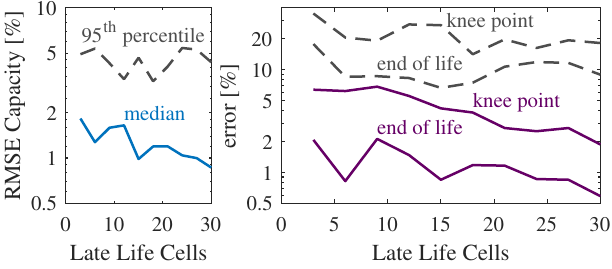}
    \caption{Comparison of the end of life and knee point errors for the limited data test. The dashed lines represent the 95\textsuperscript{th} percentiles while the solid purple lines represent the median performance across the 40 cells.}
    \label{fig:limited_data_results_comp}
\end{figure}

The results of the limited data test are presented in Fig.\ \ref{fig:limited_data_results_comp}. The median results were accurate using 15 cells worth of late life data, and were relatively consistent from training sets having as little as 6 cells with late life data. The 95\textsuperscript{th} percentiles were variable, but were around 4\%, 10\% and 20\% for the RMSE capacity, end of life percentage error and knee point percentage error respectively.

\begin{figure}
    \centering
    \includegraphics[width=.9\columnwidth]{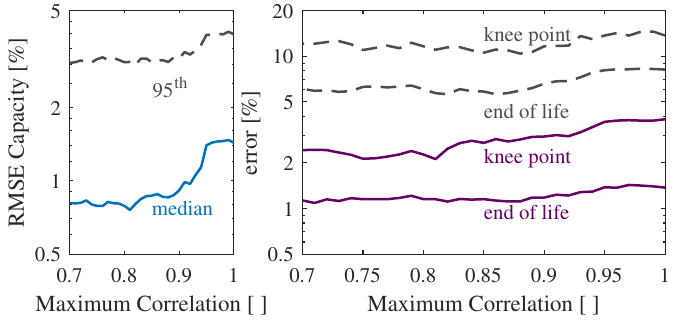}
    \caption{Comparison of the end of life and knee point errors for the varied maximum shared correlation. The dashed lines represent the 95\textsuperscript{th} percentiles while the solid purple lines represent the median performance across the 1,000 forecasts for each test point.}
    \label{fig:results_maxsims}
\end{figure}

Finally, we investigated the impact of changing the maximum shared correlation allowed between features, Fig.\ \ref{fig:results_maxsims}. Elsewhere in this work, the maximum value was 0.85, but here it was varied between 1 and 0.7. As seen, there was a distinct improvement in RMSE capacity under a value of 0.95, while a steadier improvement also occurred for end of life and knee point performance.

\section{Discussion}

The consistent selection and subsequent performance with the voltage-based features V\textsubscript{2,3} and V\textsubscript{1,2} strongly suggests that time spent in specific state of charge ranges had an impact on degradation rate. This is to be expected given the physics of battery degradation \cite{reniers2019modelsreview}, since side reactions, such as growth of the anode solid electrolyte interphase, are dependent on potential. Since the high voltage region feature V\textsubscript{3,4} and the low voltage region V\textsubscript{1,2} were found to correlate very closely with the first selected feature, V\textsubscript{2,3}, the latter was generally prioritised. The reason is that as the cells degraded, since voltage limits were reached sooner, greater periods of time were spent at high and low constant voltages during charge and discharge. This in turn reduced the amount of time spent in the voltage mid-range. Therefore the mid-range voltage feature is strongly correlated with $\Delta Q$, because in a single feature it indirectly indicates increased time spent at both high and low voltages.

Temperature was never returned from the feature selection process despite it being a known factor in battery ageing. All cells were cycled in thermally controlled environments which perhaps meant that a sufficiently wide range of temperatures was not explored to elucidate the dependence of degradation on temperature. Similarly, the automated feature extraction process was found to be robust to significant reductions in data frequency, accuracy and precision in other trials. 

Overall, the feature selection approach gave accurate results in the large-scale test, with median percentage errors for end of life predictions of 1.3\%. The mean error of 2.0\% represented at least a three-fold improvement on previous work published with similar data \cite{fermin2020kneepoints,severson2019data1}. This may be due to the use of training data across the entire life of each cell, allowing for variability in usage, rather than point measurements.  

The RMSE\textsubscript{$Q$}, RMSE\textsubscript{$\Delta Q$} and PE($t$\textsubscript{EoL}) results suggest that our feature engineering approach produced successful predictions of both capacity and end of life. We speculate that because the relationships between features and capacity changes is generally monotonic and smooth, the linear correlation approach for feature selection works well. The accurate capacity forecasts led to three quarters of knee point estimates lying within a single day of the measured value and 95\% of the profiles predicted the knee point within just two and a half days of the observation.

The limited late-life data test produced evidence of the versatility of the approach, even in the face of significantly restricted training data. This makes the technique more viable in the real-world. Nevertheless, the process cannot completely remove the need for comprehensive ageing data to end of life. Data-driven approaches cannot necessarily make accurate predictions outside of the range of their training data.

The clear improvement in accuracy when the allowed maximum correlation between features was reduced shows that a diverse input is crucial for performance. Imposing this condition either avoided overfitting to the training set and/or found input features that would not otherwise be obvious.

There were still a few outlier predictions, and Fig. \ref{fig:eol_results} shows some end of life estimates  far from a perfect prediction. Using medians is a more robust measure of overall performance, compared to using mean errors, but because medians ignore the extent of outliers some caution is required \cite{murphy2012MLprobabilistic}. 

Unfortunately, where there were larger errors between forecasted and observed end of life, this was not matched by having larger credible intervals at these points. The calibration score, the proportion of capacity observations within the predicted $2\sigma$ interval \cite{richardson2019GPRtransition}, was 0.42 for the entire large-scale trial, and for end of life predictions it was up to 0.75. Both values are below the target of 0.95, strongly suggesting the model was overconfident in its uncertainty estimates. Exploring and improving credible intervals provides an interesting avenue for future work, alongside looking at the uncertainty around use prediction. The large-scale of the tests used here allowed for reliable assessment of credible intervals. Without easily accessible large battery data sets, cross-validation represents a strong option to assess uncertainty performance. 

The k-fold cross-validation approach for quantifying performance acted as an alternative measure of accuracy instead of credible intervals. The large number of results contained in Fig.\ \ref{fig:result_histogram_plots} showed an estimate of the performance for a typical prediction and a typical poor prediction of health, both useful pieces of information, and usually hidden by the use of a mean average as a summary statistic.

\section{Conclusions}

A combined feature selection and machine learning approach for battery health prediction was proposed and tested, producing knee point forecast errors of 0.45 days or 2.6\% across 600 predictions. That success was due to an accurate capacity forecast, with half of all profiles having a root mean square error of under 0.83\% capacity when predicting over full lifetimes. A further trial showed that, as might be expected, having more data generally led to better predictive performance. The results also showed that this approach is capable of handling lesser qualities and quantities of data without unduly impacting performance for end of life prediction, despite the changing degradation rates. K-fold cross-validation produced sufficient results to calculate multiple summary statistics. Medians and higher percentiles are an informative pair of measures when used in conjunction and future work should aim to use them, especially for any asymmetric error measures.

Interesting open questions remain regarding the presented procedure and data-driven approaches in general. The number of features required could be further investigated, using a wider range of datasets. Other areas of research are improved credible intervals and tackling datasets without regular health measurements. In a real-world application, the algorithm will still be applicable and be able to extract features relating to battery usage. However, regular measurements of health may not be available in the same way as for a laboratory dataset. Solving this is an open challenge that will require new approaches, for example unsupervised learning, or using a different and more general health metric to train the model.

The feature selection approach using easily understood features provides a level of insight unavailable from black box machine learning techniques. Information on how different features correlate with one another, and which features impact degradation, is extremely useful to a user alongside the accurate forecasts of capacity, end of life and the knee point.

\begin{IEEEbiography}[{\includegraphics[width=1in,height=1.25in,clip,keepaspectratio]{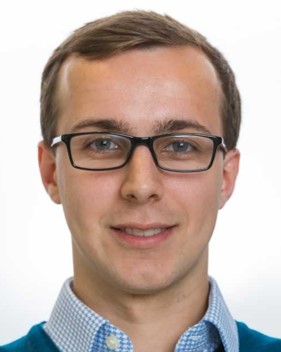}}]{Samuel Greenbank}
(M '21) is a PhD researcher within the Battery Intelligence Lab, Department of Engineering Science, University of Oxford, UK. He read for an MPhys in Physics at Lincoln College, University of Oxford, graduating in 2017. His PhD focusses on data-driven approaches to lithium-ion battery health diagnosis and prognosis.
\end{IEEEbiography}

\begin{IEEEbiography}[{\includegraphics[width=1in,height=1.25in,clip,keepaspectratio]{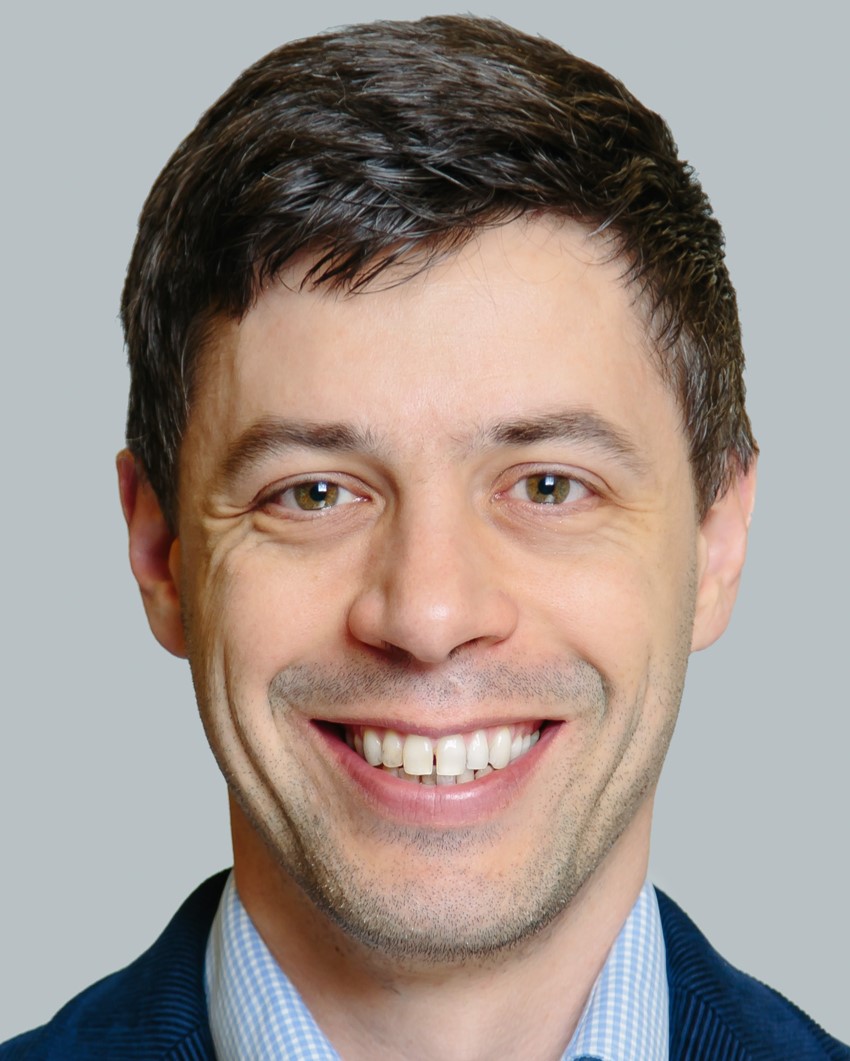}}]{David Howey}
(M’10 – SM’17) received the B.A. and M.Eng. degrees from Cambridge University, Cambridge, U.K., in 2002, and the Ph.D. degree from Imperial College London, London, U.K., in 2010. He is currently an Associate Professor at the Department of Engineering Science, University of Oxford, Oxford, U.K. His research is focused on energy storage systems, including projects on model-based battery management, degradation, thermal management, and energy management for grid storage. He is also co-founder of Brill Power Ltd., a company spun-out of his lab in 2016 focused on advanced battery management system topologies.
\end{IEEEbiography}

\end{document}